\documentclass[prc,aps,preprint,showpacs,showkeywords,,tightenlines,nofootinbib]{revtex4}
\usepackage{graphicx}
\usepackage{bm}
\usepackage{slashed}
\newcommand{\deep}{\mbox{D(e,e$^\prime$p)n}}
\newcommand{\gevcsq}{\mbox{$\rm (GeV/c)^2$}}
\newcommand{\gevc}{\mbox{$\rm GeV/c$}}

\newcommand{\exx}[1]{\mbox{$\cdot 10^{#1}$}}
\newcommand{\qsq}{\mbox{$Q^2$}}
\newcommand{\qvec}{\mbox{$\vec q$}}
\newcommand{\ppm}{\mbox{$p_{m}$}}
\newcommand{\degree}{\mbox{$^\circ$}}

\begin{document}

\title{Modern Studies of the Deuteron:\\
  from the Lab Frame to the Light Front}

\author{Werner Boeglin and Misak Sargsian}
\affiliation{Department of Physics, Florida International University, Miami, FL 33199}

\date{\today}

\begin{abstract}
We review the recent progress made in studies of deuteron structure at small internucleon distances. 
This progress is largely  facilitated by the new generation of experiments in deuteron electrodisintegration 
carried out at unprecedentedly high momentum transfer. The theoretical analysis  of these data 
confirms  the onset of the high energy eikonal regime in the scattering process which allows  one to separate 
long range nuclear effects from the effects genuinely related to the short distance structure of the deuteron.
Our conclusion is that for the first time the deuteron is probed at relative momenta beyond 300 MeV/c without dominating 
long range effects. As a result, at these large nucleon momenta the cross section 
is 
sensitive to the nuclear dynamics at sub-fermi distances.   Due to large internal momenta involved  we are dealing 
with the   relativistic bound state that is best described by the light-cone momentum distribution of nucleons in the 
deuteron.  We present the first attempt of extracting the deuteron  light-cone momentum  distribution function  from 
data and discuss  the importance of this quantity for  studies of QCD structure of the bound nucleon  
in deep inelastic scattering off the  deuteron. 
 We conclude the review giving an outlook of the next generation of high energy  experiments which will 
extend our reach to much smaller distances in the deuteron.

\end{abstract}
\pacs{21.45.Bc,25.30.-c,27.10.+h,11.80.Fv}
 
\maketitle

\section{Introduction}

The deuteron, discovered in 1931 by Harold Urey (see e.g. \cite{discovery}),  was a theoretical puzzle until the discovery of  the 
neutron in 1932.   Since then it  has been one of the best 'laboratories' for testing our understanding of the 
nuclear forces.   The evolution of our view of the deuteron was intimately related to  the advances in our understanding of 
the dynamics of the strong interaction.

As an apparent proton-neutron bound state the deuteron  was initially used to explore the large distance phenomena of 
the $pn$-interaction, which  are sensitive  to rather  general properties of  the nuclear force such as its interaction range
and the scattering length. 

However after the discovery of the deuteron's quadrupole moment, it was clear 
that the deuteron's structure  reflects  more intricate properties of  the nuclear force  such as the  tensor interaction.  
The discovery of pions indicated that the deuteron in addition to proton and neutron may contain also explicit 
pionic degrees of freedom and  
indeed  pion exchanges between proton  and neutron were soon discovered in electromagnetic reactions involving  
deuteron\cite{Hockert:1974qt,Fabian:1976ne}.

Furthermore, the discovery of a rich spectrum of baryonic resonances~({R})  implied also, in addition to 
the pn component,   the possibility for the existence of 
$NR$ and $RR$ components in the deuteron~(see e.g.  Ref.\cite{Weber:1978dh}).  
Even richer structures  have been predicted  with the advent of  Quantum Chromodynamics~(QCD) as 
a fundamental theory of strong interactions.     One of the interesting QCD predictions involving 
the deuteron is the possibility of observing  quark-gluon currents in  high energy and  momentum transfer (hard)
reactions involving the deuteron. Such a possibility is identified by a specific energy dependence of the
cross section of the hard reaction (referred as quark-counting rule)  as well as  by polarization observables following 
from the helicity conservation in  quark-gluon interactions\cite{Brodsky:1976mn}. 
Another unique QCD prediction is the possibility of hidden-color states in the deuteron  comprising  of two color-octet baryons 
resulting in a colorless deuteron\cite{Harvey:1980rva,Ji:1985ky}.  

All above mentioned  resonating and QCD properties however reside, kinematically,  beyond the NN inelastic threshold which corresponds to deuteron 
internal momenta above $\sim 370$~MeV/c. Thus this requires an ability of probing significantly shorter distances  in the deuteron 
in which case the overlap of the constituent proton and neutron can be substantial so  that one can expect the onset of  
the quark degrees of the freedom in the NN system. 

The understanding of the   short-range properties of the deuteron has important relevance  to the short range dynamics
of  nuclei and nuclear matter with recent observations of the dominance of $pn$ short-range correlations in 
nuclei\cite{Piasetzky:2006ai},  their role in a momentum  sharing in asymmetric 
nuclei\cite{Sargsian:2012sm,Hen:2014nza} as well as a possible 
$p^{-4}$ scaling of the high  momentum distribution of nucleons in nuclei\cite{Hen:2014lia}.

This review addresses the present status of the research of short-range properties of the deuteron and expected advances in  
such  studies at new high energy facilities.

The most direct way  of probing the short distance structure of the deuteron, is to probe large internal momenta 
in which case due to the large virtuality of the intermediate states the deuteron wave function is defined  
by  short distance dynamics. 

There are  several reactions which  are expected to probe  large internal  momenta in the deuteron,
such as elastic $ed\rightarrow ed$ processes at high momentum transfer~(for a review see Ref.\cite{Gilman:2001yh}), 
 or inclusive $ed\rightarrow e^\prime X$ processes 
in the quasielastic region at the so-called large $-y$ region~(for review see Ref.\cite{Arrington:2011xs}).  
These reactions  have larger cross sections  and 
historically they were the first aimed at probing short range properties of the deuteron.
However due to the specifics of these reactions only integrated properties of  the deuteron momentum distribution are
accessed.  As a result  the ability of using  these reactions to  probe the deuteron at extremely large   internal momenta is rather restricted.

The most direct way of  probing the internal structure of the deuteron  is to study  the exclusive 
\begin{equation}
e + d \rightarrow e^\prime +  p + n
\label{reaction}
\end{equation}
reaction in which one of the nucleons is struck by  the incoming electron and the other is a spectator to the 
reaction. 

 First such exclusive experiments were attempted in the early 60's using 
 electron probes of a few 100 MeV.  Due to the very small duty factor of these early accelerators and 
the extremely small cross section of the coincidence reaction only measurements at small nucleon momenta 
were possible\cite{Croissiaux:1962zza,Bounin:1965zz}. 

Starting in the 70s, when the duty factor of electron accelerators increased, attempts to probe larger nucleon momenta 
were carried out providing first cross section data for nucleon momenta up to 
300 MeV/c~\cite{Antufev:1975ei, Antufev:1974sy} and in the 80s cross section measurements at internal 
nucleon momenta beyond 300 MeV/c were carried out at SACLAY\cite{Bussiere:1981mv,TurckChieze:1984fy}. 
The theoretical analysis  of these data established that due to the lack of resolving power of the probe
($Q^2 \ll M_N^2$~GeV$^2$)  
the cross section  of the reaction (\ref{reaction}) at  internal relative $pn$  momenta $\ge 300$~MeV/c  
was dominated by long range processes such as final state interaction, intermediate isobar contributions 
as well as meson exchange currents.

The situation changed only after  the appearance of  the high energy and high duty cycle CEBAF accelerator,  were it was,
for the first time,  possible to perform   dedicated large momentum transfer ($Q^2 > M_N^2$~GeV$^2$)  experiments 
of the reaction (\ref{reaction}).

 In this paper we review the chronology of the experimental studies of exclusive 
 electro-disintegration of the deuteron (\ref{reaction}) and elaborate
 how the increase of the transferred momentum made it possible to  access  and probe the smaller internal 
 distances in the deuteron.  We demonstrate the qualitative changes that high momentum transfer processes bring 
 into the observables and the emergence of the new theoretical approximation in describing the reaction (\ref{reaction})  in 
 the high energy limit. 

 One of the  main changes  in the high energy  limit   is the emergence of the light-front dynamics 
 in probing the relativistic internal structure of the deuteron and the possibility of extracting  the
 light-cone momentum distribution of the bound nucleon in the deuteron.   We present the first such extraction 
 using  the recent high momentum transfer  data from 
 Jefferson Lab. 
 
 We conclude the review demonstrating  how the eikonal regime allows an extension 
 of the deuteron studies to
 internal momenta  in and above the GeV region.  Such momenta  can be experimentally studied at Jefferson Lab after the 
 12~GeV energy upgrade.

\section{The Modern View of the Deuteron}
\label{sec2}

We start with decomposing  the deuteron state  vector into the Fock states restricted by the total spin, S=1 and isospin, T=0 quantum numbers of  the deuteron:
\begin{equation}
\Psi_d = \Psi_{pn} + \Psi_{\Delta\Delta} + \Psi_{NN^*} + \Psi_{hc} +  \Psi_{NN\pi} \cdots
\label{Fstates}
\end{equation}
the "$\cdots$"'s include the contributions from higher Fock components and higher mass constituents.  In following we will discuss 
separately each component presented in Eq.(\ref{Fstates}).

\medskip

\noindent {\bf \boldmath $pn$ Component:} Kinematically one expects pionic degrees of freedom to become relevant 
at deuteron internal  momenta exceeding ${\sqrt{s_{thr}-4m_N^2}\over 2}\approx 370$~MeV/c. However 
the empirical evidence suggests that the $pn$ component is dominant for deuteron internal momenta up to $650$~MeV/c (see e.g. 
discussion in Ref.\cite{Frankfurt:2008zv}). This can be understood based on the following facts: 
(i)  the proportionality of the $N\pi N$ vertex to  the pion momentum, 
(ii) the form factor of $N\rightarrow \pi N$  transition being hard $\sim \exp{\lambda t}$ with $\lambda \ge 3$~GeV$^{-2}$
\cite{Frankfurt:1988nt} and
 (iii) the processes in which  the high energy probe couples to the exchanged pion in the deuteron 
is significantly suppressed at high momentum transfer.
 These facts  indicate that the dominant inelastic component can be the $\Delta\Delta$ rather than 
the $N\pi N$ component which will extend 
the $pn$ dominance  for up to $p\sim 800$~MeV/c. Despite the possibility for such a dominance, 
the $pn$ component of the deuteron wave function currently is reliably understood  
only for up to $400$~MeV/c. Many factors contribute to this uncertainty such as the  insensitivity of the small distance phenomena 
to the accuracy of the NN phase shifts, relativistic effects and the treatment of the pion threshold in the NN channel.

\medskip

\noindent {\bf \boldmath $\Delta\Delta$ Component:} 
Due to the large cross section of the $\pi N\rightarrow \Delta$ transition and the above discussed suppression of the 
 $N\pi N$ transition  one expects the largest non-nucleonic component  in the deuteron to be the $\Delta\Delta$ component.   
The current experimental constraints on the overall contribution of the $\Delta\Delta$ component is $\le 1$\%.

\medskip

\noindent {\bf \boldmath $NN^*$ Component:} 
In principle the quantum numbers of the deuteron allow the $NN^*$ component which will correspond to the radial excitation of 
one of the nucleons in the deuteron. Such an excitation will require an energy of $\sim   600$~MeV/c corresponding 
to internal momenta similar to the $\Delta\Delta$ component ($\sim 800$~MeV/c).  However, empirically one expects a smaller
$NN\rightarrow NN^*$ amplitude  which may result 
in a contribution smaller than the contribution from  the $\Delta\Delta$ component.  Currently there is no experimental 
evidence  or constraint on the possible mixture of the  $NN^*$  component.

\medskip

\noindent {\bf \boldmath Hiden Color Component:} One of the unique predictions of QCD is the 
existence of the hidden color component in the deuteron wave function.   
In fact the color decomposition of a 6q system predicts almost $80$\% of the wave function strength to be due to 
the hidden color component\cite{Harvey:1980rva,Ji:1985ky}.  However  one expects such a dominance  only to occur 
at very large excitation energies of the NN system  when the sum of the possible two-baryonic states   in the deuteron is replaced 
by the six-quark representation.  Since such large excitation energies are relevant to the nuclear core it rises 
an interesting possibility of the $NN$ repulsive core being the result of the suppressed overlap between a hidden color dominated
configurations and the $NN$ component.
\medskip

\noindent {\bf \boldmath ${NN\pi}$ Component:} Finally, the most dominant three-particle  Fock component of the deuteron 
is  the $NN\pi$ component which one may expect to appear at excitation energies close to the pion threshold 
(corresponding to  an internal momentum of $\sim 370$~MeV/c) and 
to be sensitive to the external probe at low momentum transfer. There is  plenty of  evidence of this 
component from low and intermediate energy 
reactions which probed meson exchange currents that start to dominate at the missing momentum range of $\sim 350$~MeV/c 
consistent with the  above estimate of the pion threshold (see for example Refs.\cite{Hockert:1974qt,Fabian:1976ne}).

\section{Basic Diagrams and Kinematic  Definitions of the Exclusive Reaction}
\label{sec_diagrams}

For definiteness we assume that the detected particles at the final state  of the reaction (\ref{reaction}) are 
the scattered electron and  proton while the neutron is reconstructed through the energy-momentum conservation.
We define the transferred four-momentum as $(q_0, {\bf q})$ with a virtuality of $Q^2 = {\bf q}^2 - q_0^2$ and ($E_p,\bf p_p$) 
and  ($E_n,\bf p_n)$ as the four-momenta of  the final proton and neutron respectively.
We also define the missing momentum, $p_m$,
\begin{equation}
-{\bf p_m} \equiv {\bf p_i} = {\bf p_p} - {\bf q} = - {\bf p_n}
\label{p_i}
\end{equation}
which in the special case (see below) can be interpreted as a negative initial momentum vector 
of the bound proton which interacts with the electron.  

\begin{figure}[h]
\centering\includegraphics[scale=0.9]{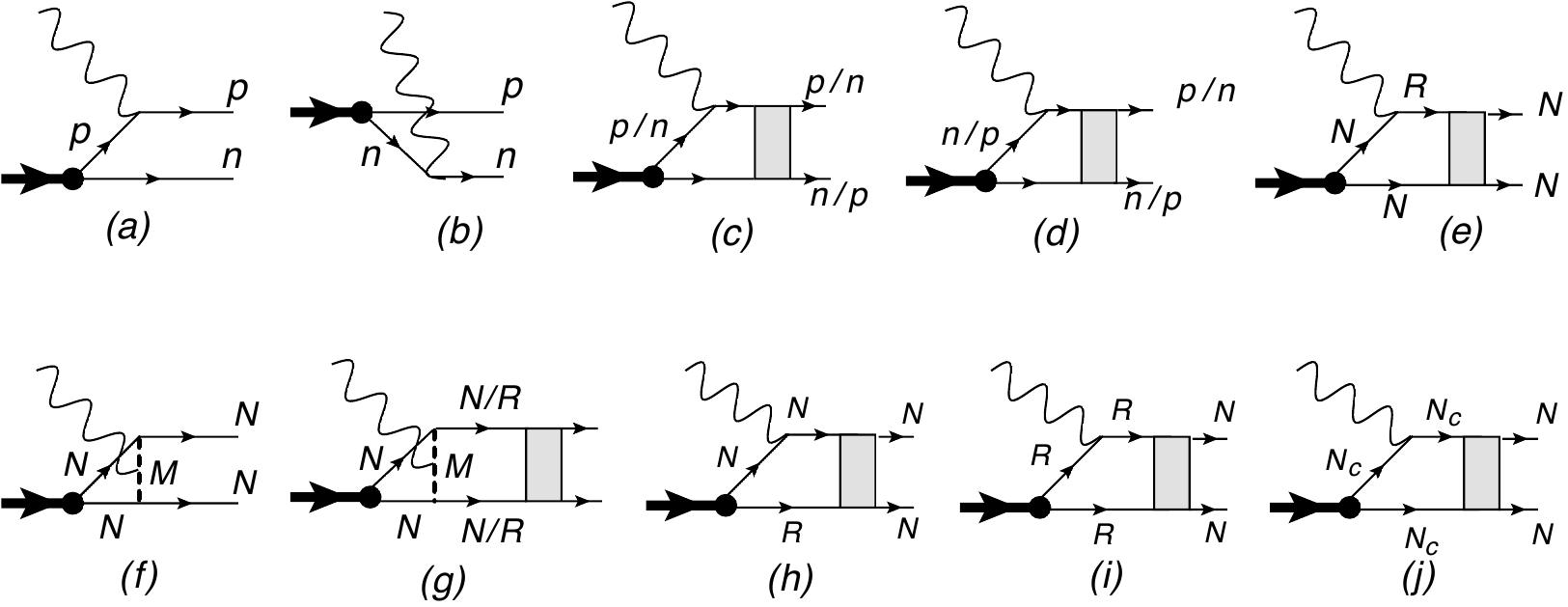}
\caption{Diagrams contributing to the exclusive $d(e,e'p)n$ reaction.}
\label{edepn_diagrams}
\end{figure}

Within the one-photon exchange approximation  the basic Feynman diagrams that describe the exclusive process 
of Eq.(\ref{reaction})   are given in  Fig.\ref{edepn_diagrams}. 
 
These diagrams can be categorized as follows:
\begin{itemize}
\item[(a)] {\bf Direct PWIA contribution:} We call the contribution of Fig.\ref{edepn_diagrams}(a)  a  direct plane-wave 
impulse approximation~(PWIA) term, in which  the detected  nucleon (proton in the example) is knocked-out  by 
the virtual photon  while the undetected  nucleon (neutron) is a spectator to the $\gamma^* p_{bound}\rightarrow p_{final}$ 
scattering. No final state interaction~(FSI) is considered and therefore final nucleons  emerge as  plane waves.
 
\item[(b)] {\bf Spectator  PWIA contribution:} In Fig.\ref{edepn_diagrams}(b) it is the  undetected  neutron which is struck 
by the virtual photon while the detected proton emerges as a spectator. Again no FSI is considered between the emerging nucleons.

\item[({c})] {\bf Direct FSI contribution:} In this case (Fig.\ref{edepn_diagrams}({c}))  the  struck proton rescatters off the spectator 
neutron and is detected in  the final state.

\item[(d)] {\bf Charge-Interchange FSI contribution:} In this case (Fig.\ref{edepn_diagrams}({d}))  
the  struck-nucleon undergoes a  charge interchange  interaction with the spectator nucleon.

\item[(e)] {\bf Intermediate State Resonance Production:} In Fig.\ref{edepn_diagrams}(e), the electromagnetic interaction 
excites the nucleon into a resonance state  which then rescatters with the spectator nucleon into 
the final proton and neutron.
 
\item[(f)] {\bf Meson Exchange Contributions:} In Fig.\ref{edepn_diagrams}(f)(g), the electromagnetic interaction 
takes places with the mesons which are exchanged between initial nucleons in the deuteron.

\item[(g)] {\bf Non-Nucleonic Contributions:} Final three terms contributing to the reaction
(Fig.\ref{edepn_diagrams}(h) (i) and (j)) are sensitive  to the non-nucleonic component of the deuteron wave 
function. Here the first two represent the baryonic and the last,  the hidden-color component 
contributions which was discussed in the Sec.\ref{sec2}.
\end{itemize}

\section{Concept of Probing the High Momentum Component of the Deuteron}
\label{sec4}

To probe the high momentum NN component of the deuteron, in the diagrammatic presentation  of the scattering 
process of Eq.(\ref{reaction}),  is to isolate the contribution from the direct PWIA 
process\footnote{Hereafter refereed as PWIA process.}  (Fig.\ref{edepn_diagrams}(a)) at such momentum transfer ${\bf q}$ and final 
proton momentum ${\bf p_p}$ that the  calculated missing momentum,  according to Eq.(\ref{p_i}),   $p_m> 300$~MeV/c.

Such an isolation however requires a suppression or reliable accounting for all the remaining contributions 
(diagrams Fig.\ref{edepn_diagrams}(b)-(j)) discussed in the previous section.   
 
Our assertion is that this can be achieved if we consider the reaction (\ref{reaction}) in {\em high energy}  kinematics 
in which  the  transferred momentum $q \ge few$~GeV/c and the  virtuality of the probe, $Q^2 > 1$~GeV$^2$, with an additional 
condition that  the final nucleons (proton in our case) carries almost all the momentum of the virtual photon while 
the recoiling nucleon (neutron) is significantly  less energetic, i.e.: 
\begin{equation}
p_p \sim q \sim few~ \mbox{GeV/c}   \ \ \ \mbox{and} \ \ \  |p_n| = |\vec q - \vec p_p|  \sim few \ hunderd ~\mbox{MeV/c} 
\ \ \mbox{and} \ \ Q^2 > 1 ~ GeV^2.
\label{henc}
\end{equation}

The effects of the above conditions are different  for different diagrams of  Fig.\ref{edepn_diagrams}(b)-(j), which 
we can categorize as {\em kinematical}, {\em dynamical} and  the mixture of both.

The suppression of the  spectator PWIA diagram, Fig.\ref{edepn_diagrams}(b),  is purely {\em kinematical}, since 
in this case the amplitude of the scattering will be defined by the deuteron wave function, $\sim \psi_d(p_p)$ with 
the initial momentum of  $p_p\sim few$~GeV/c as compared to the PWIA term  which is  proportional to 
$\sim \psi_d(p_n)$ with $p_n\sim few\ hundred$~MeV/c. 

The  diagrams containing meson exchange currents (Fig.\ref{edepn_diagrams}(f),(g)) will 
be suppressed {\em dynamically}, since in the limit of $Q^2 \gg m_{meson}^2 \sim 1$GeV$^2$ they are
suppressed  compared to the PWIA term by an extra factor of $Q^{6}$\cite{Sargsian:2001ax,Sargsian:2002wc}. 
Another  dynamical suppression occurs in  the scattering followed by the charge-interchange rescattering 
(Fig.\ref{edepn_diagrams}(d). In this
case  the suppression is  due to an extra $s^{-1/2}$ factor as well as a much stronger $t$ dependence  in the $pn\rightarrow np$  
amplitude as compared to the $pn\rightarrow pn$ amplitude that enters in the direct FSI  contribution (Fig.\ref{edepn_diagrams}({c})\cite{Sargsian:2009hf}. 
The same is true for 
processes involving non-nucleonic components of the deuteron wave function~(Fig.\ref{edepn_diagrams}(h-j)), since in this case 
rescattering amplitudes are non-pomeron-exchange type decreasing  with $s$ as compared to the almost $s$-independent  
$pn\rightarrow pn$ amplitude.  Additionally one expects  negligible contributions due to non-nucleonic components for 
deuteron internal momenta up to $700-750$~MeV/c\cite{Frankfurt:2008zv}.

Finally,  the suppression of  processes involving intermediate baryonic resonance production (Fig.\ref{edepn_diagrams}(e)), which 
is expected to be large in the $\gamma N \rightarrow \Delta$ channel,  is due to both 
kinematical and dynamical reasons. Kinematically,  in  the high energy limit it is possible to probe the lower $q_0$ side of 
the quasi-elastic peak (corresponding to Bjorken variable $x_{Bj}= {Q^2\over 2 m_N q_0} > 1$) which is maximally away from the 
inelastic threshold of $\Delta$ electroproduction.  Dynamically, due to the spin-flip nature of  the $\gamma^* N \rightarrow \Delta$ transition
one expects a much steeper falloff of the transition form-factor with $Q^2$ as compared to the elastic 
$\gamma^* N \rightarrow N$ scattering\cite{Stoler:1993yk,Ungaro:2006df}.

The above discussion leaves us with the dominating contributions from the PWIA  (Fig.\ref{edepn_diagrams}(a)) and 
direct FSI  (Fig.\ref{edepn_diagrams}({c})\footnote{Referred hereafter as FSI diagram.} diagrams only.  
There is no obvious reason for the suppression of the  latter diagram, 
since in the high energy limit the  amplitude of  $pn\rightarrow pn$ rescattering  at small angles  is dominated by 
the pomeron-exchange and is practically energy independent.
However the most important change of the character of FSI in the high energy limit is the onset of the 
eikonal regime, in which case FSIs exhibit a strong angular anisotropy\footnote{This should be 
contrasted with the  large and almost isotropic FSI at low and intermediate energies.} being large at transverse  
and small at longitudinal directions of the recoil (slow) nucleon production.

Thus,  we expect that in  the high energy limit,  defined according to Eq.(\ref{henc}),  the cross section of 
the process  (\ref{reaction}) will be defined mainly  by the PWIA~(Fig.\ref{edepn_diagrams}(a))  and 
FSI~(Fig.\ref{edepn_diagrams}(c))  processes.   In this case the concept of probing the high momentum 
component of the deuteron is related to measuring  process (\ref{reaction}) at large values of $p_m$ which 
in the lab frame of the deuteron  within PWIA  corresponds to a large internal momentum in the 
deuteron.  Such large missing momenta should be measured at forward or backward recoil nucleon angles 
which will minimize FSI effects allowing the  direct access  to  the PWIA term.

\section{Previous  $\deep$ Experiments and the Impossibility of Probing the Deuteron at Short Distances}
\label{sec5}

Before discussing the recent high energy experiments we briefly review the previous experiments at low
and intermediate energies.

Since the beginning of electron scattering experiments in the 1950's it was recognized
 that coincidence experiments  have the potential to provide a new, detailed information
 on the nuclear wave function~\cite{Durand:1959zz}. 
 One of the first d(e,e'p)n experiments was performed 
 at the  Stanford Mark III accelerator, where the coincidence cross section was 
 measured for a very low missing momentum~\cite{Croissiaux:1962zza} at a 
 four-momentum transfer, $Q^2= 0.085$ $\gevcsq$. Shortly thereafter the range in 
 missing momenta probed was extended to almost 100 MeV/c by the experiment 
 of  P. Bounin and M. Croissiaux at the Orsay Linear Accelerator~\cite{Bounin:1965zz}. The smallest 
 coincidence cross sections accessible in these early experiments were determined by 
 the small duty cycle of the available accelerators (of the order of 1\exx{-5}). 
 For a given average current, a small duty cycle  leads to very high instantaneous
 particle rates and large contributions of accidental coincidences which makes the measurement of 
 small cross secession very difficult.
 
 Starting in the 1970s a  new generation of experiments  were performed at intermediate electron beam energies and 
 higher duty factors, further extending the measured range of missing momenta.
 Experiments carried out  at  the Kharkov Institute with the 2 GeV linear accelerator (duty cycle $\approx 2\exx{-4}$) 
 were able to extend the missing momenta probed to 300 MeV/c and found large deviations 
 from the PWIA results for missing momenta above 100 MeV/c~\cite{Antufev:1975ei,Antufev:1974sy}. 
 These measurements were followed by experiments
 at Saclay\cite{Bussiere:1981mv,TurckChieze:1984fy} in the 1980s that extended the missing momentum coverage for the first
 time to 500 MeV/c using new instrumentation and the Saclay linear accelerator (ALS) with the duty factor of $1\exx{-2}$.
 
 \begin{figure}[!ht]
\centering\includegraphics[width=1.0\textwidth,clip=true]{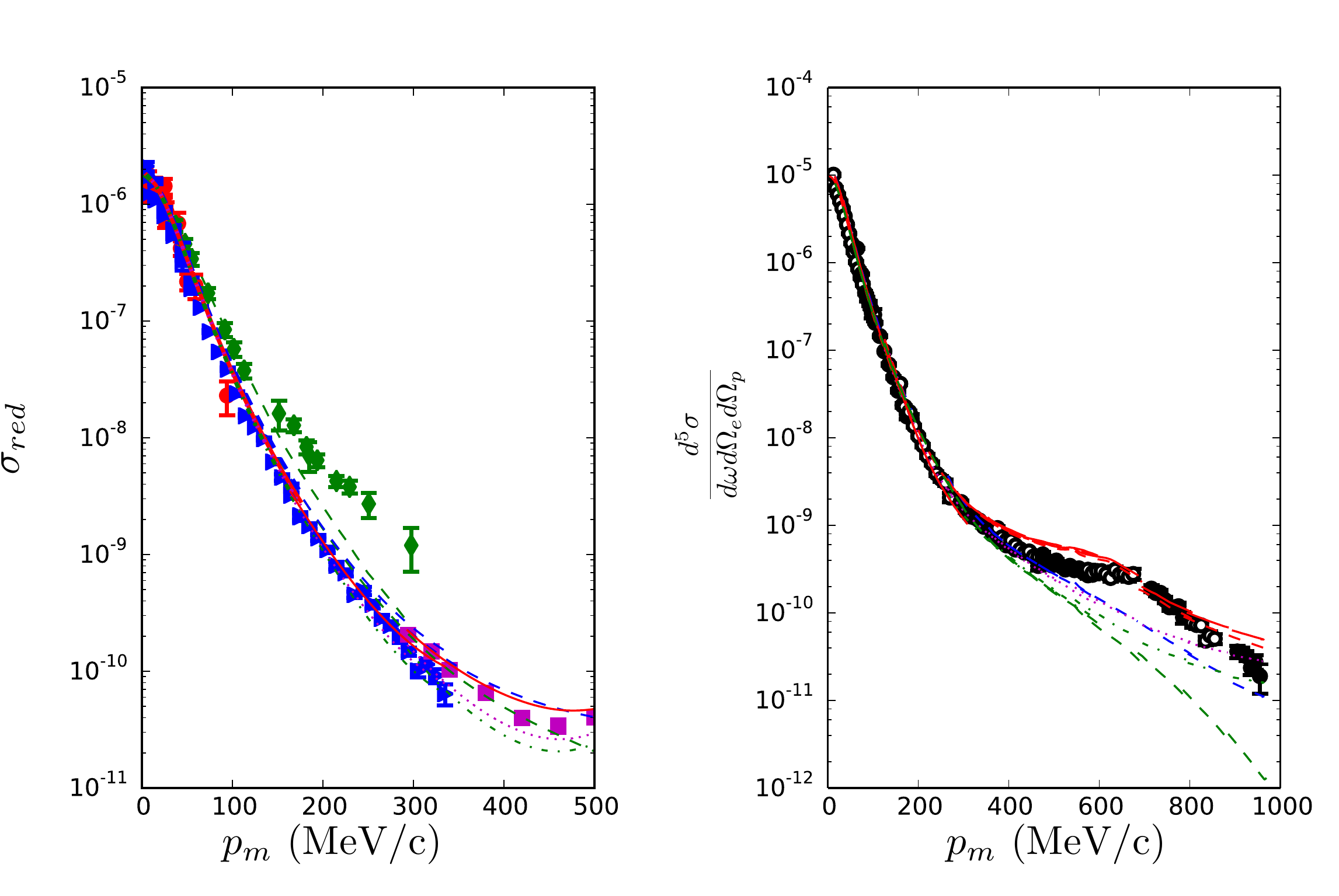}
\caption{ left: The reduced cross sections as measured at Saclay together with the earlier experiments. The red circles are from ref.~\cite{Bounin:1965zz}, 
the green diamonds from Ref.~\cite{Antufev:1974sy}, the blue triangles from Ref.~\cite{Bussiere:1981mv} and the magenta squares from 
Ref.~\cite{TurckChieze:1984fy}. 
The lines are calculations by H. Arenh\"ovel~\cite{Are13}. The green dashed line is a PWIA calculation, the blue dashed line corresponds to PWIA, 
the green dash-dotted line includes FSI, the dotted magenta line includes FSI+MEC, the solid red line includes FSI+MEC+IC and the dashed red 
line includes FSI+MEC+IC+R (R: relativistic corrections)  
Right: The cross sections measured in the Mainz experiment~\cite{Blomqvist:1998fr} compared to calculations by H. Arenh\"ovel. 
The labeling of the curves in the two panels is identical.}
\label{fig:saclay_mainz}
\end{figure} 

 The large missing momentum data in these experiments were taken at a quite small momentum transfer. Interestingly, 
 the measured cross sections at large $p_m \sim 350$~MeV/c   turned out to be quite close to the PWIA prediction. 
The  calculations\cite{Fabian:1976ne} including direct and spectator PWIA as well as FSI, meson exchange currents (MEC) and 
intermediate $\Delta$ resonance contribution (IC) demonstrated that at these 
specific kinematics a cancellation seems to occur between  FSI, MEC and IC contributions resulting in a calculated 
momentum distribution which is in quite good agreement with the PWIA prediction which includes both direct and 
spectator contributions.

With the advent of new  high duty factor ($\sim 1$)  electron accelerators in the 1990's  it became possible to measure very small cross
 sections with very little accidentals in a comparatively short amount of time. 
 This made it possible to extend the measurement of the  $\deep$ cross section up to missing momenta of 928 MeV/c 
 in a deuteron electro-disintegration experiment at the Mainz Microtron (MAMI)~\cite{Blomqvist:1998fr,Boeglin:2008pw}. 
 However in order to reach these extreme kinematic settings with the available electron beam energy and detection capabilities, virtually 
 no room was left to select the kinematics in such a way as to minimize FSI, MEC and IC contributions. 
 Indeed, it was found that the cross sections at high missing momenta were dominated by 
 long range processes such as FSI, MEC and IC contributions. The comparison with the theory demonstrated that for missing momenta above $\approx 300$~MeV/c the large FSI,  MEC and IC contributions do not mutually cancel and completely obscure
 the short-range PWIA contribution (Fig.~\ref{fig:saclay_mainz}).
    
The dominance of IC and MEC contributions were understood based on the fact 
that at $\ge 300$~MeV/c one is very close to the inelastic threshold in which the scattering proceeds through the 
copious pion and isobar production in the intermediate state of the  reaction.
 The large and isotropic FSI is  due to the fact that at  moderate momentum transfer $q$, the struck and  recoil nucleons 
in the final state  have small and comparable  momenta at $p_m\ge 300$~MeV/c. Consequently the 
rescattering is taking place at small relative momenta at which  two nucleons are 
in the relative $S$-state and have a  large interaction cross section. The dominance of the relative $S$-state leads to the 
observed  isotropic distribution of the FSI strength.

\section{First High $Q^2$ $\deep$ Experiments }
\label{sec6}
 
 With the availability of the Continuous Electron Beam Accelerator Facility~(CEBAF) at Jefferson Lab (JLab),  new  experiments
 became possible with a beam energy of up to 6 GeV,  a duty factor of one and a total beam current of up to 200$\mu A$. 
This machine, unique in the world,   allowed one for the first time
to study the $\deep$ reaction at both high $\qsq$($\ge 1$)~GeV$^2$ and high missing momenta. 
  
  \medskip
  \noindent {\bf Experiments:} 
Two recent JLab $\deep$ experiments, using the Cebaf Large Acceptance Spectrometer (CLAS) in Hall B and the two high 
resolution spectrometers (HRS) in Hall A,  measured the $\deep$ cross sections for 
large missing momenta, $p_m$  up to $550$ MeV/c at various $\qsq$  ($> 1$~GeV$^2$)
and neutron recoil angles, $\theta_{nq}$ (the angle between the 
recoiling neutron and the  three-momentum transfer $\qvec$). 
The CLAS experiment~\cite{Egiyan:2007qj} took advantage of the large 
acceptance of its detector making it possible to measure a wide range of kinematic settings simultaneously. 
The $\deep$ cross section was measured for a range of four-momentum transfers $ 1.75 < \qsq < 5.5 $ $\gevcsq$, 
for each $Q^2$ setting centered around $\qsq = 2 \pm 0.25, 3 \pm 0.5, 4\pm 0.5, 5 \pm 0.5$~GeV$^2$.
The cross sections were measured as a  function of $\theta_{nq}$ for three ranges of missing momenta: 
(i) $\ppm < 100$ MeV/c  (ii) $200 < \ppm < 300$ MeV/c and (iii) $400 < \ppm < 600$. 
In addition, the extracted cross sections were integrated over the range of azimuthal angles,  $\phi$ (the angle between 
the electron scattering plane, $(eq)$ and the reaction plane defined by the momentum transfer vector and the momentum 
vector of the ejected proton, $(qp_p)$) covered by the CLAS acceptance. 
This experiment provided a very nice overview of the general behavior of the $\deep$ cross section over a wide  kinematic range.

In the Hall A experiment the two HRS were  used to  measure the  neutron recoil angle, $\theta_{nq}$, dependence of 
the $\deep$ cross sections  for  fixed sets of $Q^2$  and missing momenta, $p_m$.
In contrast to the CLAS experiment, in the Hall A experiment the kinematic range probed by each spectrometer setting was 
much narrower. This allowed the determination of  a consistently  averaged kinematics for each measured points.
This in turn simplifies the comparison of the experimental data with theoretical calculations as there is no need to perform a 
 full Monte-Carlo  calculations to include the acceptance effects into the theory  prediction.
 In the Hall A experiment the kinematic settings were optimized to measure angular distributions for the missing momenta 
$\ppm = 200\pm 20, 400 \pm 20, 500 \pm 20$ MeV/c. 
The largest measured value of $\theta_{nq}$ was determined by the largest proton momentum accessible to the proton spectrometer. 
Angular distributions were extracted for the four-momentum transfers $\qsq = 0.8 \pm  0.25, 2.1\pm  0.25, 3.5\pm  0.25$~GeV$^2$. 
First results obtained at the highest four-momentum transfer $\qsq = 3.5$ $\gevcsq$ have been published\cite{Boeglin:2011mt}. 

\medskip

\noindent{\bf Data:} The  CLAS/Hall B  experiment measured the recoil neutron's  angular  and momentum dependences of the  
cross section in which the data were integrated over the large intervals of 
neutron momenta (for the $\theta_{nq}$ dependence) and the whole range of recoil angles for the $p_n$ dependence. 
The theoretical 
comparison with the data required an identical  integration  of the theoretical cross section 
over the  acceptance  of CLAS detector.

The Hall A data too  measured the recoil neutron's angular and momentum dependences but for significantly smaller kinematical 
bins.  As a result the theoretical predictions could be directly compared with the data.  For the angular dependence the data were 
presented in the form of the ratio:
\begin{equation}
R = {\sigma^{exp}\over \sigma^{PWIA}},
\label{Rexp_pwia}
\end{equation}
where $\sigma^{exp}$ represents the differential cross section, ${d\sigma\over dE^\prime_e d\Omega_e d\Omega_n}$,
and $\sigma^{PWIA}$ is the same differential cross section calculated within PWIA.  The reason of considering such a ratio
was that in the high energy limit it should approach  unity for diminishing FSI contribution. Thus the  ratio, $R$,  is most 
appropriate for studying FSI effect in the $ed\rightarrow e^\prime pn$ reaction.

For studying the  momentum distribution the  Hall A  data were  presented in the form of the reduced cross sections, 
defined as 
\begin{equation}
\sigma_{red}  = {\sigma^{exp}\over K\sigma_{ep}}
\label{red}
\end{equation}
where  $\sigma_{ep}$ is the theoretical cross section of the  electron scattering off the bound proton  
with momentum $p_i$, in which  the co-factor, $K$, is   fixed in such  a way that  in the PWIA limit $\sigma_{red}$ 
gives the deuteron momentum 
distribution $n_d(p_i)$. 

It is worth mentioning that theoretical calculations of  $\sigma^{PWIA}$  (in Eq.(\ref{Rexp_pwia}))
and $\sigma_{ep}$ (in Eq.(\ref{red}))  contain uncertainties due to relativistic effects and  off-shellness of the 
bound proton which increase  with  increasing  neutron recoil momentum. 
Therefore the precise theoretical analysis 
will require a careful account of these effects.

 As it will be discussed in Sec.\ref{sec7}
the angular distributions of both experiments clearly exhibit  the onset of the eikonal regime for the FSI processes.
The momentum distributions however differ in these experiments.  The CLAS data for 
the given missing momentum  bin  have been integrated over the whole range of the recoil nucleon's angle. 
As a result the momentum distributions do not separate  regions in which FSI contributions are minimal to 
allow the extraction of the experimental deuteron momentum distribution at large missing momenta.

On the other hand the much smaller kinematic 
bin sizes in the Hall A data enabled the extraction of momentum distributions for fixed recoil angles with well defined kinematic 
settings for each $\ppm$ bin. As a result 
the dependence of the  reduced cross section is  studied as a function of the recoil neutron angle 
and thus for various degrees of FSI contributions. 
As will be shown in Sec.\ref{sec8} by  selecting  a  $\theta_{nq}$  region in which FSI is minimal  
one can  measure  the  $\deep$ cross  sections  being dominated 
by the short range structure of the deuteron.

\section{Window to  Short Distances in the Deuteron at Large Momentum Transfer Processes}
\label{sec7}

As mentioned previously with increasing momentum and energy transfers,  the eikonal regime
is expected to be established. 
This leads to a strong  angular anisotropy of FSI, 
with  the $pn$ reinteraction dominating  at kinematics in which the recoiling nucleon is produced at almost transverse angles 
with respect to the transfered momentum $q$.

However the most important feature of the eikonal regime of rescatterings  is that it creates a 
unique   possibility  for the  cancellation of FSIs at large values of missing (or recoil)  momenta. To see this 
we present the $ed\rightarrow epn$ scattering amplitude within the generalized eikonal approximation 
(GEA)\cite{Sargsian:2001ax,Frankfurt:1996xx,Sargsian:2009hf}:
\begin{eqnarray}
A^\mu  && = A_0^\mu + A_1^\mu = \sqrt{2(2\pi)^3 2E_r} \Psi_d(p_i)j^\mu_N(p_i+q,p_i) -  \nonumber \\
& &  {\sqrt{2(2\pi)^3} \over 2}
\int {d^3p^\prime_i\over (2\pi)^3} {\sqrt{2E^\prime_r} \sqrt{s(s-4m_N^2)}\over 2E^\prime_r q}
 {f_{pn}(p^\prime_i-p_i)\over p_{i,z} + \Delta - p^\prime_{i,z} + i\varepsilon} j^\mu_N(p^\prime_i + q, p^\prime_i)\psi_d(p^\prime_i),    
\label{ample}
\end{eqnarray}
where $p_r\equiv p_n$ is the recoil nucleon momentum and $\vec p_i = -\vec p_r$ is the initial momentum of the struck nucleon 
within PWIA. The initial momentum of the struck nucleon in the FSI amplitude is defined as $p^\prime_i \ne - \vec p_r$.
In the high energy limit the  factor, $\Delta \approx {q_0\over {\bf q}}(E_r - m_N)$, where $E_r$ is the energy of the 
recoil nucleon. Also $s = (q+p_d)^2$ is the total invariant energy of the reaction.
 
 The Eq.(\ref{ample}) allows one to make several  general conclusions about the structure of FSI in the high energy limit.
 
 \noindent
 {\bf First - Kinematical Constraints:}  The existence of the 
 pole in the rescattering amplitude of Eq.(\ref{ample})  indicates that the rescattering defines the initial longitudinal momentum of the 
 struck nucleon at,
 \begin{equation}
 p^\prime_{i,z} = p_{i,z} + \Delta.
 \end{equation}
 Because both $\Delta$ and $p_{i,z} = - p_{r,z}$  are defined by the  external kinematics of the reaction one can have 
 a  way of controlling   the magnitude of the initial nucleon momentum before the rescattering thus suppressing FSI
 as compared to the PWIA amplitude.
 For example if one satisfies the condition:
 \begin{equation}
|p^\prime_{i,z}| > |p_{i,z}|,
\end{equation}
then the rescattering amplitude will be  suppressed due to the larger momentum 
of the deuteron wave function entering in the FSI  term  as compared to that of the PWIA term.   
Since $\Delta$ is always positive, such a condition is automatically satisfied for 
positive $p_{i,z}$'s which correspond to the production of the spectator nucleon in backward  directions. 
 
One can also estimate the initial  transverse momentum of the struck nucleon in the FSI amplitude 
 noticing  that in the high energy limit  the $pn$ rescattering has a  diffractive nature:
 \begin{equation}
f_{pn} = \sigma_{tot}(i+\beta)e^{{B_{NN}\over 2}t} \approx  i\sigma_{tot}e^{-{B_{NN}\over 2} k^2_t},  \ \ \beta \ll 1,
\label{diffpn}
 \end{equation}
 where  $\sigma_{tot}$ is the total cross section of $pn$ scattering and $B_{NN}$ is 
the slope factor of  the elastic differential cross section.  Using this and Eq.(\ref{ample}) we notice that the  transverse momentum
 of the initial nucleon is related to the measured transverse momentum $\vec p_{i,t}  = - \vec p_{r,t}$ as:
 \begin{equation}
  \vec p^{ \ \prime}_{i,t} = \vec p_{i,t} + \vec k_t,
  \label{transk}
\end{equation}  
where $<k^2_t> \sim {2\over B_{NN}}$. This relation indicates that FSI will be maximal at $\vec p_{i,t}\approx - k_t$  corresponding 
to smaller $p^\prime_{i,t}$   and minimal  at $|\vec p_{i,t}|\ll  |k_t|$  corresponding to $p^\prime_{i,t}\sim k_t$, which 
enters the deuteron wave function in the FSI amplitude.
 
The  above observations on  longitudinal and transverse components of the struck nucleon's initial momenta 
allow us to predict that   the FSI  will be small at small $p_r$  and  
very anisotropic  at large $p_r$: dominating at transverse ($p_r\perp q$) and being suppressed  at 
parallel/antiparallel ($p_r|| q$)  kinematics.    

\noindent
{\bf Second - Cancelation of FSI:} Using  the fact that in the high energy limit  the $pn$ scattering amplitude  
is predominantly imaginary (Eq.(\ref{diffpn})), 
one  can demonstrate that at large recoil momenta $p_r$, it is always possible to find kinematics 
in which the FSI amplitude  being large is cancelled in the cross section.  The reason for such a cancelation 
is in the fact that the PWIA and  imaginary  part of the FSI amplitude enter with opposite signs 
creating an option in which case the PWIA/FSI interference cancels the  modulus-square of the FSI amplitude.   

To demonstrate this, on a qualitative level,  we consider the expression of  $|A^\mu|^2$ from  Eq.(\ref{ample}), performing 
the following simplifications which preserve the  eikonal nature of the 
scattering: (1) we  use the analytic form of 
Eq.(\ref{diffpn}) considering  the $pn$ amplitude as purely  imaginary,   (2) we neglect by small principal value part of 
Eq.(\ref{ample}) and factorize the electromagnetic current $j^\mu$ from the integrand.  After these approximations one 
obtains  from Eq.(\ref{ample}) the following condition for the cancelation of the FSI contribution in the cross section:
\begin{equation}
\psi_d(p_r) = {1\over 8}\int {d^2k_t\over (2\pi)^2} \sigma_{tot} e^{-{B_{NN}\over 2}k_t^2 - {B_{NN}\over 2}\Delta^2} 
\psi_d(\tilde p_r),
\label{fsican}
\end{equation}
where $\tilde p_r = (p_{r,z} - \Delta, p_{r,t} - k_t)$ is defined by the pole value of the integral in Eq.(\ref{ample}).

\begin{figure}[!ht]
\centering\includegraphics[width=0.96\textwidth,clip=true]{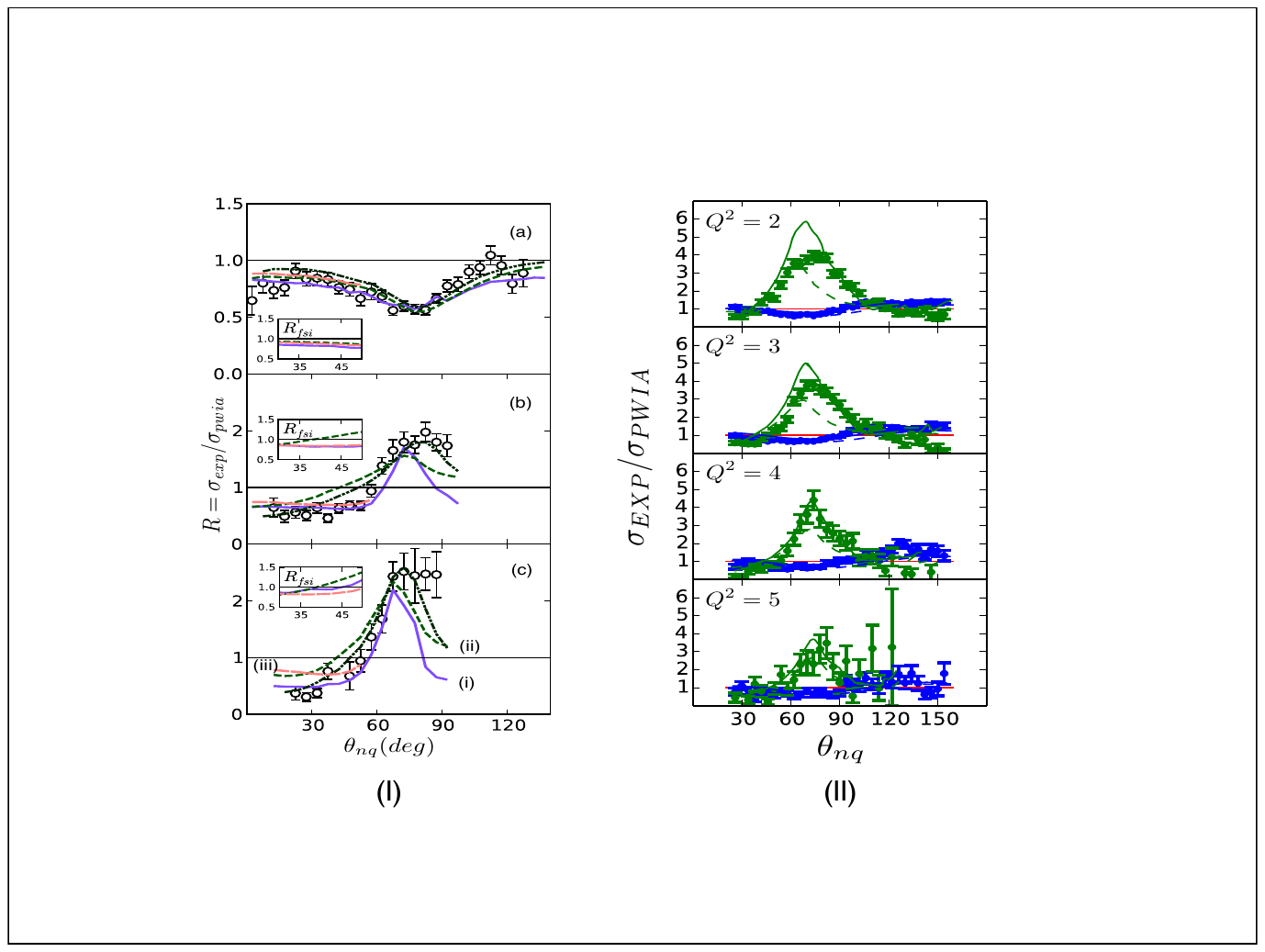}
\caption{ The ratio $R(\theta_{nq})$ of Eq.(\ref{Rexp_pwia}). (I) Hall A data  (a) $\ppm = 0.2$ $\gevc$, (b) $\ppm =
  0.4$ $\gevc$ and (c) $\ppm = 0.5$ $\gevc$.  Solid (purple) curves (i) prediction of  
  Ref.\cite{Frankfurt:1996xx,Sargsian:2001ax,Sargsian:2009hf}  
  using the  CD-Bonn potential, short dashed (green) curves (ii) prediction of Ref.\cite{Laget:2004sm},   
  dashed-double-dot curves   prediction of Ref.\cite{Laget:2004sm} including MEC and IC effects, and
  long-dashed (orange) lines (iii) prediction of Ref.\cite{Jeschonnek:2008zg}. 
  Insets: $R$ for  the range of $35\degree \leq \theta_{nq} \leq45\degree$.
  (II) CLAS/Hall B data. The data points that exhibit  rescattering peaks at transverse kinematics  correspond to 
  $400 \le \ppm  \le 600$~MeV/c, and the others without rescattering  peaks  correspond to  $200 \le \ppm  \le 300$~MeV/c.
  Solid curves - prediction of Ref. \cite{Laget:2004sm} and dashed - prediction of Ref.\cite{Sargsian:2009hf}.
   }
\label{fig:R_angular}
\end{figure} 

We can now solve the above equation analytically if we simplify the deuteron wave function in the form 
of $\psi_d = C e^{-\alpha_d p^2}$, in which  by using a Harmonic-Oscilliator~(HO) model of the deuteron  one relates 
the $\alpha_d$ parameter  to the rms radius and momentum of the 
deuteron: $\alpha_d = {r_{rms}\over 2 p_{rms}}$. Using this approximation  and neglecting 
the recoil factor $\Delta$ one arrives at  a very simple analytic relation for the cancellation of the FSI:
\begin{equation}
p_{r,t}^2 = {1\over \alpha_d} \ln{{32\pi \alpha_d\over \sigma_{NN}}}\approx {1\over \alpha_d} \ln{\alpha_d},
\end{equation}
where in the derivation we also used the relation $\alpha_d\gg B_{NN}$. In the last part of the equation we used   the known 
value of $\sigma_{pp} = 40$~mb, which results in ${32\pi\over \sigma_{NN}}\approx 1$~GeV$^2$. 
For the estimation of $\alpha_d$ according to the  HO approximation we use $r_{rms}^d = 2.13$~Fm\cite{Sick:1996plb} and 
$p_{rms}=135$~MeV/c\footnote {Which is estimated using the deuteron momentum  distribution 
based on V18 NN Potential\cite{Wiringa:1994wb}.}  obtaining  $\alpha_d \approx 40$~GeV$^{-2}$ and 
\begin{equation}
|p_{r,t}| \approx 300~\mbox{MeV/c}. 
\end{equation}
This result indicates  that there is always a characteristic transverse momentum defined by the properties of the deuteron 
and the characteristic  spatial range of   the high energy $pn$ scattering ($\sigma_{pn} \sim r_{pn}^2$) 
at which the  interference between PWIA and FSI amplitudes (screening term) cancels 
the modulus-square  of the FSI amplitude  (rescattering term).
Such cancelation  happens at two angles corresponding to the production of the recoil nucleon in the forward and backward directions,
namely:
\begin{equation}
\sin(\theta_{c1}) =  \sin({\pi\over 2}+\theta_{c2})  = {p_{r,t}\over p_r} \sim {300~\mbox{MeV/c} \over p_r}.
\end{equation}
This simplified estimate shows that starting at $p_r> \sim 300$~MeV/c at two angles of spectator nucleon production 
one expects a cancellation of the FSI effects thus opening a window for  probing deuteron at large internal momenta.

It is worth mentioning  that this cancellation is inherent to high energy  scattering  and  is observed in  all theoretical 
calculations based on the eikonal approximation 
(see e.g. \cite{Frankfurt:1994kt,Bianconi:1994bx,Frankfurt:1996uz,Frankfurt:1996xx,Laget:1998tp,
Jeschonnek:2000nh,CiofidegliAtti:2000xj,CiofidegliAtti:2004jg,Laget:2004sm,
Sargsian:2001ax,Sargsian:2004tz,Sargsian:2005ru,Sargsian:2009hf,Cosyn:2010ux,Ford:2013zca}).

The above discussed feature of FSI has been experimentally confirmed  in the  first  high 
$Q^2$($\ge 2$~GeV$^2$) measurements at Jefferson lab\cite{Egiyan:2007qj,Boeglin:2011mt} discussed in Sec.\ref{sec6}.
In  Fig.\ref{fig:R_angular} the ratio (\ref{Rexp_pwia}) is presented as a function of spectator nucleon angle, $\theta_{nq}$, for 
different values of recoil neutron (or missing) momenta.  All data are at sufficiently large $Q^2$ kinematics  for 
which  one expects  the onset of eikonal regime.
Fig.\ref{fig:R_angular}~(I) which presents the  Hall A data\cite{Boeglin:2011mt}, shows  a clear diffractive pattern of 
the FSI in which at small    recoil momenta ($p_r< 300$~MeV/c)  the ratio, $R$,
is depleted due to  screening effects ($R<1$)  while 
at larger momenta ($p_r> 300$~MeV/c) one observes a  diffractive FSI peak with $R>1$.  
The same feature are visible in Fig.\ref{fig:R_angular}~(II) where CLAS/Hall B
data\cite{Egiyan:2007qj} are presented.  In this case despite the large range of integrations of recoil neutron momenta 
one still can see distinctive  diffractive pattern of FSI. 

The other important feature of Fig.\ref{fig:R_angular}  is the observation of the  above discussed cancellations of the 
FSI in  forward $\sim 40^0$ and backward $\sim 120^0$ directions of recoil neutron production.  
These cancelations allow one to probe the "genuine" deuteron momentum distribution beyond $300$~MeV/c.   In further discussions we will concentrate only on 
the forward direction of the recoil neutron production.  The reason is that the kinematics of the backward production of a recoil neutron 
 is close to the inelastic threshold of 
$\gamma^*p$  and as a result,  for not very large $Q^2$  ($2 \le Q^2 \le 4 $~GeV$^2$) there is 
a considerable contribution from  processes with intermediate $\Delta$ resonance production (Fig.1(i)).

\section{First Measurements of the Deuteron Momentum Distribution Beyond 300 MeV/C}
\label{sec8}

The discussions of the sections (\ref{sec4}) and (\ref{sec7})   allow us to make two main  conclusions: (i) providing 
kinematic constraints of Eq.(\ref{henc})  allows us to suppress all but FSI contributions which obscure the scattering 
from the high momentum component of the deuteron; 
(ii) high energy eikonal kinematics  create a unique condition for the cancellation of the FSI contribution in 
the  forward, $\theta_{nq}\le 40^0$,  direction of spectator nucleon production. 

As it was discussed in Secs.(\ref{sec6}) and  (\ref{sec7})  the first two high $Q^2$  experiments performed recently  at 
Jefferson Lab\cite{Egiyan:2007qj,Boeglin:2011mt}
satisfy these   conditions  and  their measured spectator nucleon angular distributions clearly  demonstrate the onset of 
the eikonal regime for FSI, Fig.\ref{fig:R_angular}.  Thus one expects that the momentum dependence of the cross section of 
the reaction (\ref{reaction}) measured at $\theta_{nq}\le 40^0$ directions  and $p_r > 300$~MeV/c  momenta   will be related to the genuine 
high momentum component of the deuteron wave function.

Such a dependence was extracted from the experimental data of Ref.\cite{Boeglin:2011mt} in which the reduced cross section (\ref{red}) was measured 
for fixed $Q^2=3.5$~GeV$^2$ for recoil nucleon momenta up to $550$~MeV/c.  The ability of this experiment to separate narrow 
$\theta_{nq}$ bins in the momentum distribution was important for separating regions with different degrees of FSI contribution.

\begin{figure}[!ht]
\centering\includegraphics[height=0.4\textheight,clip=true]{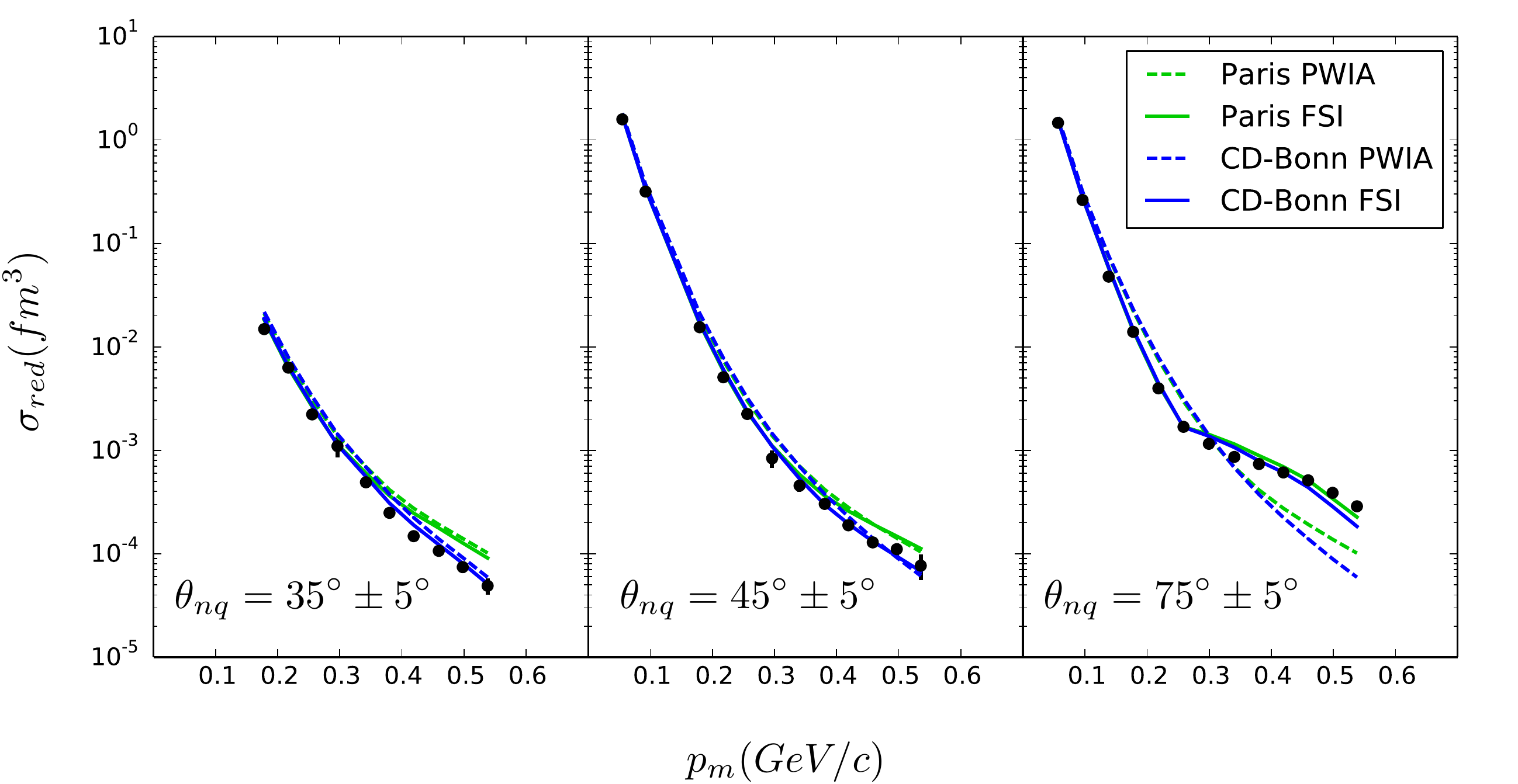}
\caption{ The experimental reduced cross sections (momentum distributions) for three values of the recoil angle $\theta_{nq}$. 
The solid lines indicate calculations\cite{Sargsian:2009hf}  including FSI and the dashed ones are PWIA calculations.}
\label{fig:pm_dist}
\end{figure} 

As an illustration in Fig.\ref{fig:pm_dist}  the missing momentum dependence for fixed recoil~(neutron) angles 
$\theta_{nq} = 35$, $45$ and $75$ degrees are shown for the  reduced cross section defined in Eq.(\ref{red}).
For comparison, theoretical  calculations  of Ref.\cite{Sargsian:2009hf} with and without FSI are also shown. 
For the large recoil angle, ($\theta_{nq}=75^0$)  corresponding to an electron kinematics of $x_B \approx 1$ FSI dominates 
for missing momenta above 0.3 GeV/c and the sensitivity to the choice of the deuteron wave function  is greatly reduced. 
In fact the momentum dependence in this case strongly resembles that of  the low and intermediate energies (Fig.\ref{fig:saclay_mainz}).

The momentum dependence of the reduced cross section at large $p_{m}$ qualitatively changes at small angles.
As  Fig.\ref{fig:pm_dist} shows for $\theta_{nq} = 35^0$ and $45^0$  FSI contributes significantly  less and the 
reduced cross section is much more sensitive to the details of the high momentum distribution, 
allowing us for the first time to discriminate between different models of the deuteron wave function.

\section{The First Extraction of the Light Cone Momentum Distribution of the Deuteron}
\label{sec9}

The availability of   high $Q^2$ data on exclusive  electrodisintegration of the deuteron  provides,  for the first time,   
the  possibility to extract directly the light cone momentum distribution of the deuteron $\rho_d(\alpha)$, where $\alpha$ 
represents light-cone momentum 
fraction of the struck nucleon normalized  in  such way that the total light-cone momentum fraction of the deuteron is 2. 
The $\alpha$ variable satisfies the condition: $0 \le \alpha \le 2$ and from  the momentum-energy conservation it follows
that $\alpha+\alpha_r = 2$ where $\alpha_r$ is the momentum fraction of the recoil nucleon.

The function $\rho_d(\alpha)$ plays an essential role 
in studies of  inclusive and semi-inclusive deep inelastic scattering (DIS) off the deuteron which are used for 
studies of the partonic structure of the  neutron.    For example, for inclusive scattering,  in the high $Q^2$ limit, the experimental 
DIS structure function of the deuteron $F_{2d}$ can be expressed through the bound nucleon  structure functions $F_{2N}^{bound}$ as follows:
\begin{equation}
F_{2d}(x_{Bj},Q^2) = \int\limits_{x_{Bj}}^2   F^{bound}_{2p}({x_{Bj}\over \alpha},Q^2) \rho_d(\alpha) {d\alpha\over \alpha} + 
\int\limits_{x_{Bj}}^2   F^{bound}_{2n}({x_{Bj}\over \alpha},Q^2) \rho_d(\alpha) {d\alpha\over \alpha}, 
\label{dis}
\end{equation}
which indicates that the knowledge of the structure function of the bound proton $\tilde F^{bound}_{2p}$  as
well as $\rho_d(\alpha)$ is necessary for the extraction of $F^{bound}_{2n}(x_{Bj})$ which is largely unknown at $x_{Bj}>0.7$.
Here the function $\rho_d(\alpha)$ satisfied two sum rules:
\begin{equation}
\int\limits_0^2 \rho_d(\alpha) {d\alpha\over \alpha} = 1 \ \ \ \mbox{and} \ \ \ \int\limits_0^2 \rho_d(\alpha) \alpha {d\alpha\over \alpha} = 1,
\label{norms}
\end{equation}
where the first follows from the requirement of the conservation of the baryonic number and the second represents 
the momentum sum rule (see e.g. \cite{Frankfurt:1981mk}). 

It is worth noting that ${\rho_d(\alpha)\over \alpha}$ is analogous to the partonic 
distribution function $f_i(x)$ but for nucleonic degrees of freedom (for details see e.g. Ref.\cite{Freese:2014zda}).  
And similar to partonic distribution functions one can introduce 
the unintegrated function $\rho_d(\alpha,p_t)$ such that
\begin{equation}
\rho_d(\alpha) = \int \rho_d(\alpha,p_t) d^2 p_t.
\label{rhoa}
\end{equation}

 \begin{figure}[!ht]
\centering\includegraphics[height=0.4\textheight,clip=true]{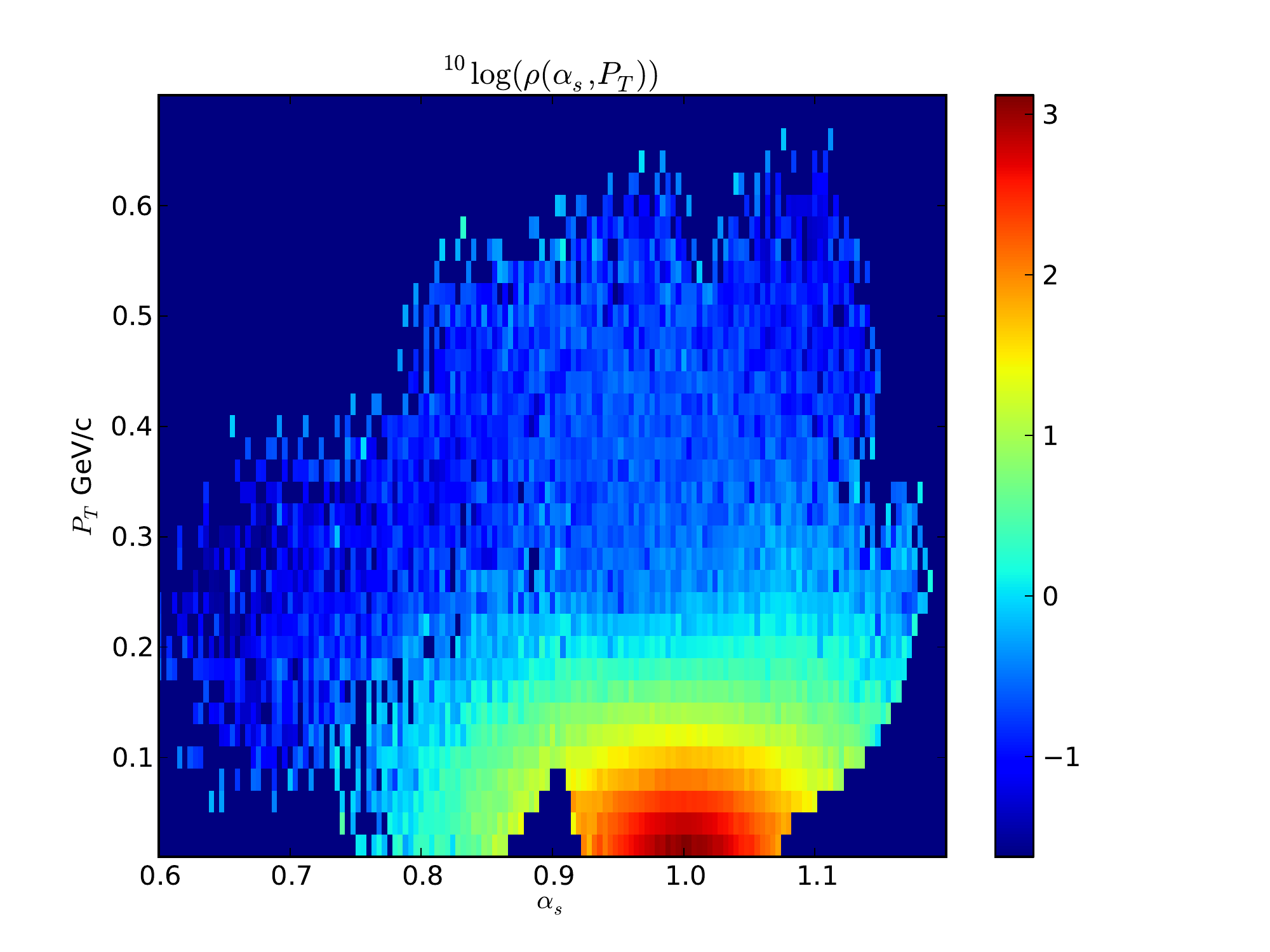}
\caption{ The experimental $\rho(\alpha, p_t)$ distribution}
\label{fig:alpha_pt}
\end{figure} 

The unintegrated density function $\rho_d(\alpha,p_t)$ can be expressed  through the light-cone wave function of the deuteron as follows:
\begin{equation}
\rho(\alpha,p_t) = {|\psi_{d}^{LC}(\alpha,p_t)|^2\over 2-\alpha}.
\end{equation}
On the fundamental level to calculate the light-cone wave function of the deuteron one needs to 
identify  the $NN$ interaction potentials on the light-front  and solve   
the Weinberg-type equations\cite{Weinberg:1966jm} for two-body bound systems.  Such a program is tremendously 
difficult (see e.g. \cite{Cooke:2001kz})  and in reality different theoretical approximations are made to  model 
the light-cone wave function of the deuteron  (see e.g. \cite{Frankfurt:1981mk}).  

One concludes from the above discussion that   the possibility of an experimental 
extraction of the $\rho_d$ function is important from both theoretical and experimental point of views. 
For theory, it will  make it possible  to check different approaches in calculating the deuteron wave function 
within light-cone dynamics and for experiment, the extracted distribution can be used as an input in studies 
of DIS  processes involving deuteron.

The idea that one can extract the $\rho_d(\alpha)$ from the reaction (\ref{reaction}) follows from the theoretical 
observation\cite{Sargsian:2009hf} that  in the high energy limit FSI processes (Fig.\ref{edepn_diagrams}({c})) do not modify the 
initial light-cone momentum fraction of the struck nucleon $\alpha$.  This can be seen analytically  by observing that the 
FSI diagram of Eq.(\ref{ample}) on the light-cone can be represented as follows\cite{Sargsian:2009hf}:
\begin{eqnarray}
 A_1^\mu & =   &  \sqrt{2(2\pi)^3 2m_N} 
\int {d\alpha^\prime d2p^\prime_{it}\over 2\alpha^\prime (2\pi)^3} {\sqrt{s(s-4m_N^2)}\over  2m_N q}
 {f_{pn}(p^\prime_{it}-p_{it})\over \alpha^\prime - \alpha - {Q^2\over 2 q^2}{E_n-m_N\over m_N}  + i\varepsilon} \nonumber \\
& &  j^\mu_N(\alpha^\prime, p^\prime_{it} + q, \alpha_i, p^\prime_{it})\psi^{LC}_d(\alpha^\prime, p^\prime_{it}),    
\label{amplc}
\end{eqnarray}
from which one observes  that in the limit of ${Q^2\over 2 q^2}{E_n-m_N\over m_N}  \ll 1$ the pole value in  the integrand 
corresponds to $\alpha = \alpha^\prime$ which indicates that the  FSI does not change  the 
light-cone momentum fraction, $\alpha$ of the  initial nucleon.    

The above result qualitatively means that the FSI will redistribute the scattering strength only in the transverse, $p_t$ direction, 
therefore if one extracts the $p_t$ integrated light-cone density function $\rho_d(\alpha)$ from the experiment, 
it will be minimally modified due to FSI.

Recently, the first attempt has been made to extract the $\rho(\alpha)$ function from the  $Q^2 = 3.5$~Gev$^2$ data measured 
at Hall A/ JLAB.  For this,  the data set  described in Sec.\ref{sec6},  
have been analyzed in terms of the light-cone variables $\alpha, p_t$.  
On an event by event basis the recoiling neutron momentum fraction $\alpha_n$  and the transverse momentum,
$p_{tn}$, (with respect 
to the  ${\bf q}$) have been calculated.  The  initial light-cone momentum fraction and the transverse momentum 
of the  interacting nucleon (proton) was calculated using the relation: $\alpha = 2-\alpha_n$ and $p_t = - p_{tn}$. 
Then in a completely analogous way as the standard $\deep$ analysis, the experimental yields for the different kinematic 
settings have been normalized, radiatively corrected and for each $\alpha,p_t$ bin a differential cross section has been 
determined. The various kinematic settings have then been combined by averaging overlapping bins.  

The extraction of $\rho(\alpha,p_t)$ was made based on the theoretical framework of light-cone PWIA  according to which:
\begin{equation}
{d\sigma\over dEe^\prime d\Omega_e d\Omega_s} = K\sigma^{LC}_{eN}(\alpha,p_t)\rho(\alpha,p_t),
\label{lcpwia}
\end{equation}
where $\alpha = 2 - \alpha_n$,  $\alpha_n = { E_n - p_{n,z} \over M_d/2}$ and $k$ is a kinematic factor.

\begin{figure}[!ht]
\centering\includegraphics[height=0.4\textheight,clip=true]{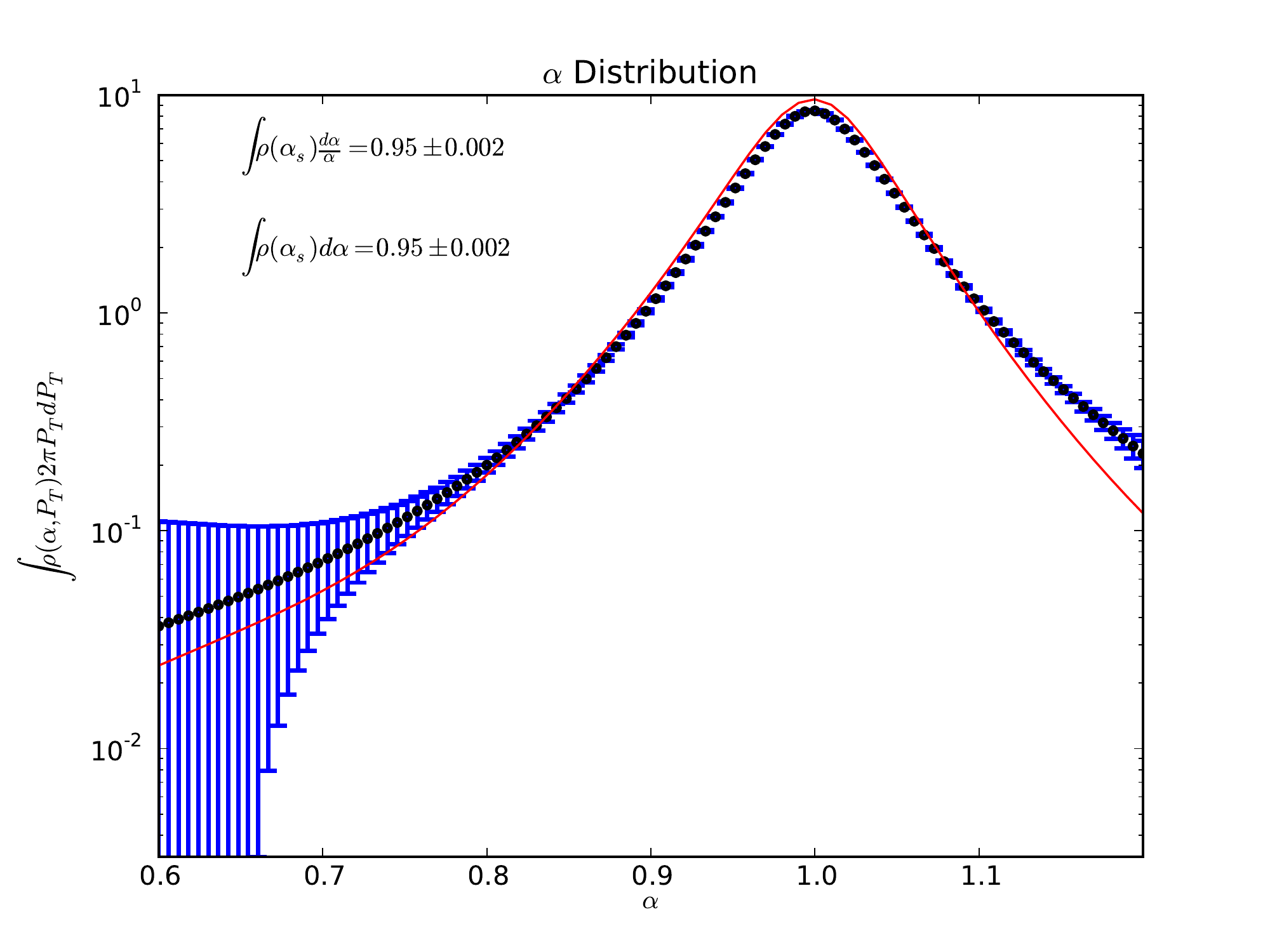}
\caption{ The experimental $\rho(\alpha)$ distribution, integrated over $p_t$}
\label{fig:rho_alpha}
\end{figure}

To extract the $\rho$ function the evaluated experimental cross sections have  been divided by the 
corresponding light cone e-p cross section,  $\sigma^{LC}_{eN}(\alpha,p_t)$,   and the kinematic factor $K$.  This, 
according to Eq.(\ref{lcpwia}) allowed  one to obtain the 'experimental' distribution $\rho(\alpha, p_t)$ presented in 
Fig.~\ref{fig:alpha_pt}.

The data presented in  Fig.~\ref{fig:alpha_pt} makes it possible to study the 
transverse momentum, $p_t$  distribution for constant $\alpha$  or vice versa.
Especially interesting is the integration of $\rho(\alpha,p_t)$ over $p_t$ for which one can check the sum rules of 
Eq.(\ref{norms}). The satisfaction of these sum rules will be indicative of the cancellation of the FSI In the $\rho(\alpha)$
distribution.

To perform the $p_t$ integration and  extract the $\rho(\alpha)$ we  fitted  model distributions to the data for slices of 
constant $p_t$, which 
was necessary as the kinematic coverage in the available data set is not complete. Using the fitted  
$\alpha$ distributions we could interpolate between bins where no data exist and use these in the integration 
over $p_t$. This procedure  then leads to the  experimental $\rho(\alpha)$ distribution, shown in Fig.~\ref{fig:rho_alpha}.    
In the same figure the results of the  baryonic and momentum sum rules are presented which 
indicate a reasonable agreement with  Eq.(\ref{norms}).  The 
agreement  with both sum rules is the  strongest indication that FSI redistributes the cross section in the transverse momentum
space  and as a result the integrated $\rho_d(\alpha)$ distribution represents the unaltered light-cone momentum 
fraction distribution of the nucleon in the deuteron. 
 
Our conclusion is that the first analysis demonstrates the feasibility of an experimental 
extraction of the integrated light-cone momentum  distribution of the deuteron and  new data with wider kinematical coverage 
will allow  to increase significantly  the accuracy of such an extraction.

 \section{Summary and Outlook for Future Studies of the Deuteron}

In this review we presented the current status of studies of the deuteron at large internal momenta and 
demonstrated that the  possibility of performing exclusive $d(e,e'N)N$ experiments  at large momentum and energy transfers 
satisfying the kinematical conditions of Eq.(\ref{henc}), opens up a window to probing the deuteron structure at small 
distances.

As we discussed, the  attempts to determine the deuteron momentum distribution for missing momenta above $300$~MeV/c 
at low momentum transfer experiments suffered from 
the  large contributions from  FSI, MEC and IC effects which are dominated 
by the long range structure of the deuteron.  We then demonstrated that by increasing the momentum and energy transfers above 
the $GeV$ limit  allows one to suppress MEC and IC contributions while for FSI the eikonal regime is established.
The eikonal regime is characterized by a very anisotropic angular distribution of the recoil nucleon production and 
it makes it possible to identify kinematical regions where FSI are mostly cancelled. 

The first high energy data confirm the angular distribution characteristic of  FSI  in the eikonal regime. 
Then,  using 
the high precision $Q^2=3.5$~GeV$^2$ data at the kinematics were one expects FSI cancelation to occur,   we 
probed the deuteron in a missing momentum region of $300-500$~MeV/c without large contributions of FSI, MEC or IC.

Another new venue  in  probing the deuteron at small distances and  high momentum transfers is the use of the approximate 
conservation of the light-cone momentum fraction $\alpha$ in the high energy limit to extract the light-cone momentum distribution 
function $\rho(\alpha)$ which is minimally altered by  FSI.

The above  concepts can be extended  to the deuteron studies at the completely 
unexplored  internal momentum  range of $\sim 1$~GeV/c.  Such studies are fascinating since for the first time one will have 
an opportunity to probe a nuclear bound state at distances relevant to the $NN$ core.

\begin{figure}[!ht]
\centering\includegraphics[width=0.6\textwidth,clip=true]{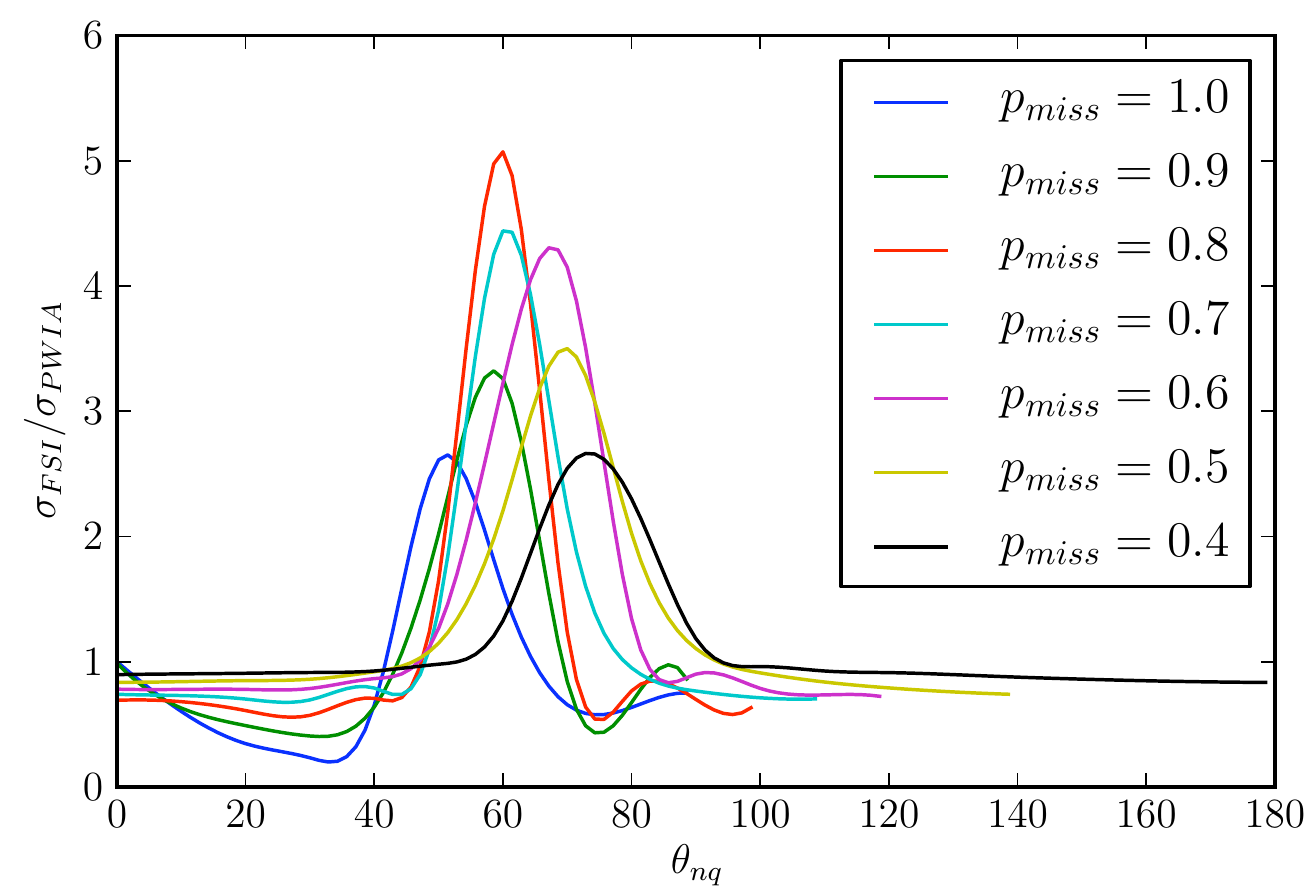}
\caption{ Ratios of the calculated cross sections including FSI effects  $\sigma^{FSI}$ to that of the
PWIA at different values of missing momenta as a function of the neutron recoil angle.}
\label{ratios}
\end{figure} 

As Fig.\ref{ratios} demonstrates, at large $Q^2$ it is still possible to identify the kinematic window of minimal FSI 
effects for missing momenta of the reaction up to $1$~GeV/c.
This gives a possibility to extend the studies of the deuteron beyond $550$~MeV/c.  A first such attempt will be made 
 in the  approved experiment of Ref.\cite{Boeglin:2014aca} which will be carried out at the Jefferson Lab
 with the electron beam upgraded to the 11~GeV.
This experiment will reach a momentum transfer of  Q$^2$ = 4.25 (Gev/c)$^2$  with recoil  
neutron  angle covering the forward,  35 - 40$\degree$,  region were one expects  reduced FSI (Fig.\ref{ratios}).

The large value of the $Q^2$ will be essential for satisfying the condition of Eq.(\ref{henc}) which is necessary for 
identifying the $1$~GeV/c neutron as a spectator (preexisting) nucleon in the deuteron, thereby providing the condition of 
probing $NN$ distances relevant to $1$~GeV/c relative momentum in the deuteron.

In addition, the possibility of the measurement of  the angular distribution for $\ppm > 550$~MeV/c will allow 
the extension of the range of $\alpha$ at which the light-cone momentum distribution of the deuteron, $\rho(\alpha)$ can be extracted.
The study of the deuteron structure on the light-cone has just started. A good kinematic coverage is required to reduce the 
model dependence of the determination of $\rho(\alpha)$. If necessary, new data could be obtained using new detector 
systems which have large acceptances allowing  high luminosity  beams. Such a situation will allow the measurement of 
very small coincidence cross sections of reaction (\ref{reaction}).  The next stage of such studies will be 
the extension of the reaction  to the case of a polarized electron beam and polarized deuteron target.
The study of the deuteron light cone wave function using polarization degrees of freedom is currently being studied.   

Finally,  the deuteron structure can be probed in deep-inelastic hard processes such as 
hard break-up of the deuteron \cite{Frankfurt:1999ik,Granados:2010cj} and deep-inelstic electro-productions of photons and  
vector mesons  aimed at  studies of the generalized partonic distributions  in the deuteron (see e.g  \cite{Cano:2003ju,Berger:2001zb}).

All the above  mentioned  studies  will be an important step in designing a new generation of experiments with 
the further goal of probing even shorter distances  relevant to 
the non-nucleonic as well as quark-gluon content of the deuteron.

\medskip
\medskip

{\bf Acknowledgments:}

The authors are  thankful to Drs.~C. Ciofi degli Atti, L.~Frankfurt, J.M.~Laget, S.~Jeschonnek, M.~Jones,
W. Van Orden, G.~Miller, M.~Strikman, E.~Voutier for helpful comments and discussions. 
This work is supported by U.S. DOE  grants under contract DE-FG02-01ER41172 and
DE-FG02-99ER41065. 




\end{document}